# The Impacts of Registration Regime Implementation on IPO Pricing Efficiency

**Qi Deng[1,2,3,*], Linhong Zheng[4], Jiaqi Peng[5], Xu Li[6], Zhong-guo Zhou[7§], Monica Hussein[7], Dingyi Chen[1†], Mick Swartz[8]**

## Abstract

We study the impacts of regime changes and related rule implementations on IPOs' initial return for China's entrepreneurial boards (ChiNext and STAR). We propose that an initial return contains the issuer's fair value and an investors' overreaction and examine their magnitudes and determinants. Our findings reveal an evolution of IPO pricing in response to the progression of regulation changes along four dimensions: 1) governing regulation regime, 2) listing day trading restrictions, 3) listing rules for issuers, and 4) participation requirements for investors. We find that the most efficient regulation regime in Chinese IPO pricing has four characteristics: 1) registration system, 2) no hard return caps nor trading curbs that restrict the initial return; 3) more specific listing rules for issuers, and 4) more stringent participation requirements for investors. In all contexts, we show that the registration regime governing the STAR IPOs offers the most efficient pricing.

Funding Source: The work was supported by Hubei University of Automotive Technology Research Fund [grant number BK202209, 2022].

†*§ Corresponding authors.

1. College of Artificial Intelligence, Hubei University of Automotive Technology, Shiyan, Hubei, China
2. School of Accounting, Economics and Finance, University of Portsmouth, Portsmouth, UK
3. Cofintelligence Financial Technology Ltd., Hong Kong and Shanghai, China
4. Department of Political Science, University College London, London, UK
5. Annenberg School for Communication and Journalism, University of Southern California, Los Angeles, California, USA
6. School of Business, Hong Kong Baptist University, Hong Kong, China
7. Department of Finance, Financial Planning, and Insurance, David Nazarian College of Business and Economics, California State University Northridge, California, USA
8. Marshall School of Business, University of Southern California, Los Angeles, California, USA

† First corresponding author, email: dy.chen@huat.edu.cn

§ Second corresponding author, email: zhong-guo.zhou@csun.edu

* Third corresponding author, email: dq@huat.edu.cn; qi.deng@cofintelligence.com

## 1. Introduction

China's capital market consists of the main boards on the Shanghai Stock Exchange (SHSE) and Shenzhen Stock Exchange (SZSE), as well as NASDAQ-like boards that cater to early-stage, fast-growth, and high-tech firms, which include the ChiNext and STAR. On February 17, 2023, the Chinese Securities Regulatory Commission (CSRC), the top government regulator for all financial markets, announced that following the success of a registration process that governs IPO listing in ChiNext and STAR, all China's boards will adopt the registration regime and move away from the decade-old approval regime. To our knowledge, there are no empirical studies among the existing literature that examine if the registration regime is more efficient in IPO pricing than the approval one. As ChiNext is the only board that has listed a substantial number of IPOs under both the approval and registration regimes, this study extends Deng et al. (2023) by comparing the pricing efficiency of ChiNext IPOs under the two different regimes. In addition, it investigates the difference of pricing efficiency under the same registration regime with two different implementations by comparing and contrasting the performance of registered ChiNext and STAR IPOs.

Since its establishment, the ChiNext board has experienced two different regulation regimes and three different sets of listing day trading restrictions. Specifically, from the launch on October 30, 2009 to December 12, 2012, ChiNext IPOs were under the approval regime with a set of listing day trading curbs; the IPOs were underpriced by an average of 34.41% (Deng and Zhou, 2015a). In an effort to improve IPO pricing efficiency, the CSRC suspended approving new IPO listing applications across all the boards from November 3, 2012 to January 17, 2014 (known as the 2013 reform). After the CSRC resumed IPO activities in January 2014, ChiNext IPOs were still governed by the approval regime and subject to a new set of listing day "hard" return caps. With a dataset up to December 31, 2017, Zhou, Hussein and Deng (2021) find that the 2013 reform actually causes even more severe underpricing, which is confirmed by Deng et al. (2023) with an expanded dataset up to June 30, 2022.[1] Thus, as an effort to revert the unintended impact of the 2013 reform and encouraged by the successful launch of the STAR board in July 2019, the CSRC extended the registration regime from STAR to ChiNext on June 12, 2020, and simultaneously lifted all the listing day return caps.[2] Deng et al. (2023) find that the IPO regime change from the approval regime to registration regime has shifted ChiNext IPO pricing from demand-driven to value-driven. Their findings suggest that even though the IPO pricing under the registration regime seems more market-oriented, the relative level of overreaction for the ChiNext IPOs under the registration regime is still higher than those under the approval regime (27.33% vs. 13.30%). This empirical result, served as the first motivation, prompts us for a future investigation on whether the registration regime indeed improves pricing efficiency for ChiNext IPOs.

As the first Chinese board that pilots a registration process for its IPOs, STAR was launched on June 13, 2019 and listed its first batch of 25 new issues on July 22, 2019. It paved the way for ChiNext and eventually the main boards to adopt the registration regime. The establishment of STAR indicates that the CSRC finally warmed up to the idea of "self-regulation." To differentiate itself from ChiNext, STAR strategically favors "hard technology" firms even though they may yet to be profitable. This strategic direction adds additional challenge in pricing the STAR IPOs. To mitigate this additional risk, the STAR board spearheaded a different implementation of the registration regime by stipulating more specific and quantifiable listing rules for the issuers and more stringent participation requirements for the investors. The rationale is that the investors with a higher level of sophistication could better evaluate the issuers with an improved access to easier-to-digest information, which would potentially reduce pricing errors and improve IPO pricing efficiency. The second motivation of this paper is to investigate if and how the registration regime could be further improved to enhance IPO pricing efficiency. The third motivation of this study is prompted by the existing variations between the ChiNext and STAR in implementing the registration regime for their IPOs; we aim to investigate whether the variations in implementations would affect IPO pricing efficiency.[3]

[1] Zhou, Hussein, and Deng (2021) provide detailed descriptions of the new rules and regulations implemented in January 2014. They coin the overhaul in Chinese stock market in 2013 as "the 2013 reform."
[2] See Deng et al. (2023) for detailed descriptions on the new rules and regulations implemented on ChiNext IPOs in 2020, to which they refer as "the 2020 reform."

The main focus of this paper is to further extend Deng et al. (2023) by thoroughly examining the variations of the two registration regime implementations and their impacts on initial return of IPOs listed on the ChiNext and STAR boards. Built upon Deng and Zhou (2015a, 2016), we analyze the two fundamental components of the IPO's initial return: an issuer's fair value, represented by an IPO's monthly return – also known as a short-term equilibrium return, and an overreaction from the investors, represented by its intramonth return, which is the difference between the monthly return and initial return. With an updated dataset, we compare the levels of initial return, fair value, and overreaction for ChiNext IPOs and STAR IPOs, as well as the determinants that drive them to understand the underlying factors causing the differences. In addition, we expand the work of Deng et al. (2023) by providing updated empirical evidence on ChiNext and STAR IPO pricing efficiency.

[3] Appendix 1 provides the details on the variations of regulation regimes (Panel A), listing day trading restrictions (Panel B), as well as the listing rules for issuer and participation requirements for investor (Panel C) for the registered ChiNext IPOs and registered STAR IPOs.

We find that all the significant variables that affect the initial return, monthly return, and intramonth return for ChiNext and STAR IPOs fall into four categories: 1) pre-listing demand, 2) post-listing demand, 3) market condition, and 4) pre-listing issuer value. We observe high similarity among the variable categories for each of the returns between the IPOs from two different boards under the registration regime, while the STAR IPOs exhibit lower relative level of overreaction. Our findings indicate that the differences between the levels and determinants of initial return, fair value, and overreaction for the ChiNext and STAR IPOs can largely be explained by the two implementation variations: 1) listing rules for issuers, and 2) participation requirements for investors.

We document that the level of relative overreaction for registered STAR IPOs is lower than that for registered ChiNext IPOs, proving that the registration regime implementation governing the STAR IPOs is more efficient in IPO pricing. We also find that the variance of relative overreaction for the registered ChiNext IPOs is lower than that for the approved ones, which provides direct empirical evidence that the registration regime produces more efficiently priced IPOs than the approval regime. Combining and summarizing the above findings, we conclude that the most efficient regulation regime in IPO pricing has the following four characteristics: 1) registration system, 2) no hard return caps and nor trading curbs that cap the initial return, 3) more specific listing rules for the issuers, and 4) more stringent participation requirements for the investors.

The potential contribution of this study is four-fold. First, it is among the early attempts to investigate the levels of relative overreaction for approved and registered ChiNext IPOs, and the results show that the registration regime produces a lower relative overreaction variance, and thus a higher pricing efficiency. The findings provide a direct and strong empirical support to the most recent decision by the CSRC of moving all China boards to the registration regime. Second, it is among the first to investigate the levels and drivers of the registered ChiNext and STAR IPOs' initial return, fair value, and overreaction, and the results indicate that the registration regime governing STAR IPOs produces lower level of relative overreaction and therefore higher pricing efficiency. Third, it documents that the most efficient regulation regime in IPO pricing depends on four specific characteristics. And last, its results have policy implications as the CSRC has been continuously searching for the best IPO process to provide the most efficient pricing.

The rest of the paper is organized as follows. Section 2 provides literature review on ChiNext and STAR IPO pricing and initial return, the background on the progression of regulation regimes governing ChiNext and STAR IPOs, as well as our research perspective and context. Section 3 covers the dataset and variable selection. Section 4 provides statistical models and empirical results. Section 5 discusses the regulation regime changes and their impacts on initial return, fair value, overreaction, and IPO pricing efficiency while Section 6 concludes the paper.

## 2. Literature Review, Regulations, and Research Context

### 2.1 Literature Review on Classic IPO Underpricing Theories

The majority of IPO underpricing literature analyzes the causes of underpricing from an information asymmetry perspective (Berle and Means, 1930; Reilly and Hatfield, 1969; Baron and Holmström, 1980; Beatty and Ritter, 1986; Rock, 1986; Welch, 1992; and Barry and Jennings, 1993). Other mainstream hypotheses include the signaling theory (Ibbotson, 1975; Bhattacharya, 1979; Welch, 1989; Michaely and Shaw, 1994; Demers and Lewellen, 2003) and from the perspective of market conditions and investor sentiment (Ljungqvist et al., 2006).

There is also a stream of more technical approaches that connects the information disclosed from an IPOs' prospectus to its initial returns. St-Pierre (2000) investigates the possibility of using prospectus information to predict an IPO's success and finds that some factors disclosed (for example, manager voting right percentage, sales growth potential and subscription price) provide meaningful indication on the initial return. Mohd-Rashid et al. (2018) find that certain selected information from prospectus (for example, the number of institutional investors, underwriter ranking and shareholder retention) signal an IPO's offer price. Loughran and McDonald (2013) investigate the tone of words in form S-1, the first document of an IPO filing to the US Securities and Exchanges Commission (SEC) and find that IPOs with higher level of uncertain text have higher initial returns.

## 2.2 ChiNext and STAR IPO Underpricing

Since both ChiNext and STAR boards cater to the early-stage high-tech issuers, most of the studies by Chinese authors focus on the potential impacts of information asymmetry due to the issuers' higher risk profiles. The degree of an issuer's research and development (R&D) activity is one of the frequently examined factors. Qiu and Wei (2021) attribute a positive correlation between R&D and IPO underpricing to information asymmetry in R&D investment. Chen and Song (2020), Zeng and Ma (2021), and Yang and Wang (2021) make similar arguments. Tang and Zhou (2023) also document a positive correlation between the IPO funds raised for R&D and initial underpricing and they attribute that to investors' expectation of a better performance in the future, using the R&D funds. Hu et al.

(2021) take the perspective of underwriters and find that experienced underwriters can reduce underpricing, as they are able to reduce the level of information asymmetry by minimizing the time gap between pricing and listing.

On the contrary, Chen and Deng (2021) report that the number of patents as a measure of R&D activity reduces the degree of IPO underpricing. Zeng and Ma (2021) point out that the positive impact of R&D investment on IPO underpricing is more apparent in high-tech and large-sized enterprises, suggesting a firm- and size-specific effect.

On the winner's curse hypothesis, Gao (2020) proposes that the time gap between the pricing and listing dates is positively related to IPO underpricing, which is in line with Wang and Liu (2016). On the contrary, Zhang, He and Tan (2021) argue that a long waiting period curbs investors' enthusiasm, which in turn causes prices to fall in the secondary market, thus reduces the severity of underpricing. The signaling theory has some support in China. Wang and Liu (2016) demonstrate that an issuer gives investors the impression of "hot sales" through underpricing in order to create more demand on new shares while Chen and Song (2020) propose that high investors' sentiment is the main reason for IPO underpricing. Zheng et al. (2017) demonstrate that an IPO's initial return and the issuer's long-term performance are negatively corelated. Yang, Zhou, and Zhou (2022) investigate ChiNext IPOs' long-term performance from an insiders' point of view and find that insiders tend to behave strategically to maintain inflated stock price in order to maximize their own level of satisfaction.

Many China-based researchers argue that regulation regime change plays an extremely significant role in Chinese IPO pricing and performance. Liu and Yu (2010) propose that the purpose of regulation is to protect legitimate interests of all participants and maintain the public's confidence in the market, and thus foster healthy development of the market. Gao (2020) finds that the IPO registration system is conducive to improve the issuer's P/E ratio but causes more severe underpricing and therefore concludes that the key factor in IPO underpricing is the institutional arrangement. Zhang and Wu (2021), on the other hand, find that the regulation-compliant follow-up investment system has a positive correlation with the degree of underpricing.

One unique aspect of China's IPO regulation is reflected through the offline (for institutional investors) and online (for retail investors) subscription rates, which measure the degree of IPO underpricing from the perspective of supply and demand. Deng and Zhou (2015a) find that the offline subscription rate is negatively related to IPO underpricing, while Chen and Song (2020) find that the online subscription rate is negatively related to IPO underpricing. Overall, scholars are still in search of different angels to explain the initial underpricing and "the most efficient" IPO pricing mechanism.

### 2.3 Research Perspective and Context

Deng et al. (2023) examine the impacts of regulation regime changes on ChiNext IPOs over the entire lifespan of the board (using the data from October 2009 until June 2022). Before the 2013 reform, the ChiNext IPOs were under the approval regime and the issuers were vetted by the regulators and therefore perceived to be of "good" quality. As a result, the investors bid for new shares aggressively without carefully examining the issuers' quality. The relative level of overreaction in that time period is 13.30%, relative to the initial return. After the 2020 reform, ChiNext IPOs are under the registration regime and the regulators only check whether the issuers' paperwork fulfills the listing rules. Thus, the investors need to examine the issuers' quality in greater detail by themselves.

The progression from an approval regime to a registration regime effectively changes the ChiNext IPO pricing process from demand-driven to value driven. Given that the general investors are ill-equipped to analyze the self-reported information by the issuers, nor are they sophisticated enough to acquire and digest private information about the issuers, the relative level of overreaction is much higher at 27.33%, relative to the initial return. They further identify three factor lines that significantly explain the impact of regulation changes on ChiNext IPOs' initial returns: 1) governing regulation regime, 2) listing day trading restrictions, and 3) issuer's profile.

This paper expands Deng et al. (2023) with an extended dataset that includes the registered ChiNext IPOs and newly available registered STAR IPOs. Compared to that for the ChiNext IPOs, the implementation of registration regime for the STAR IPOs is subject to more specific listing rules for the issuers and more stringent participation requirements for the investors. All those differences motivate us to further study their potential impacts on IPO underpricing and IPO pricing efficiency. We aim to examine the variations of the two registration regime implementations, as well as the impacts of these implementation variations on initial return, fair value, and overreaction of ChiNext and STAR IPOs.

## 3. Dataset and Empirical Methods

### 3.1 Dataset

We collect a total of 57 potential explanatory variables from Wind Financial Terminal (WFT) based on existing literature, regulatory impacts, and novelty for initial screening [4].

[4] To keep this manuscript concise, we don't report the full table of all 57 variables. However, it is available upon request, along with all the original and processed data and variables relevant to this paper. Interested readers may send an email to the corresponding authors for data access.

The variables cover a wide range of pre-listing financials and the information disclosed in IPO prospectuses on the issuers, investors, underwriters, professional intermediates, post-listing trading information, and market data. The dataset includes 394 ChiNext IPOs between August 24, 2020 and November 11, 2022 (post 2020 reform, ChiNext sample set), and 452 STAR IPOs from July 22, 2019 to August 18, 2022 (STAR sample set).[5] In addition, in order to compare the levels and variances of relative fair value and relative overreaction between the approved and registered ChiNext IPOs, we also include 349 ChiNext IPOs from October 30, 2009 to December 31, 2012 that are under the approved regime (ChiNext Approved sample set or simply Approved sample set) in our analysis.[6]

### 3.2 Return Variables and Descriptive Statistics

The initial return (*IR*), monthly return (*MR*), and intramonth return (*IMR*) are defined as follows:

$$IR_i = \left(\frac{1CP_i - OP_i}{OP_i}\right) \times 100\% \quad (1)$$

$$MR_i = \left(\frac{21CP_i - OP_i}{OP_i}\right) \times 100\% \quad (2)$$

$$IMR_i = MR_i - IR_i \quad (3)$$

where $1CP_i$ is IPO i's 1st trading day (listing day) closing price, $21CP_i$ is its 21st trading day closing price, and $OP_i$ is IPO i's offer price. The overreaction is the difference between the 21st day closing price return and initial return. All returns are measured in percentages.

In addition, we define the Relative Fair Value (*RFV*) as the relative level of monthly return as a percentage of the initial return, and the Relative Overreaction (*RO*) as the negative relative level of intramonth return as a percentage of the initial return, assuming that the 21st day closing price represents a short-term equilibrium price (Deng and Zhou 2016).[7] The $RFV$ and $RO$ for IPO$_i$ are defined below:

$$RFV_i = \frac{MR_i}{IR_i} \quad (4)$$

$$RO_i = -\frac{IMR_i}{IR_i} \quad (5)$$

[5] While the CSRC Decree #167 and the SZSE Bulletin #[2020]515 were released on June 12, 2020, they became effective for ChiNext IPOs listed after August 24, 2020 (inclusive).

[6] We don't analyze 349 ChiNext IPOs in the ChiNext Approved sample between October 30, 2009 and December 31, 2012 in this paper as they have been thoroughly examined by Deng and Zhou (2015a, 2016). We just refer their results in Table 2. We do, however, analyze the registered ChiNext IPOs because we use an extended dataset of the IPOs (the data lasts until November 11, 2022) that is longer than the dataset used in Deng et al. (2023) (with the data lasting until June 30, 2022).

Equations 4 and 5 imply that $RFV_i + RO_i = 1$.

Figure 1 provides a visual illustration of the distributions of the three return variables (*IR*, *MR* and *IMR*) while Figure 2 illustrates the distributions of the relative fair value (*RFV*) and relative overreaction (*RO*) variables over time for the 393 ChiNext IPOs (after removing one extreme outlier), along with the basic summary statistics. Similarly, Figures 3 and 4 illustrate the distributions of the same five variables for the 451 STAR IPOs over time (after removing one extreme outlier).

Table 1 provides detailed descriptive statistics for the three return and two pricing efficiency variables for the ChiNext and STAR IPOs over their entire sample periods. In panel A and for the registered ChiNext IPOs, we find that the average initial return is 164.79% with a standard deviation of 194.41%. The median return is 104.91%. It seems that the distribution is skewed to the right, presumably caused by several extremely large and positive initial returns. The average monthly return is 121.06% with a standard deviation of 169.24%. The difference between the monthly return and initial return defined as intramonth return (overreaction) is -43.73% with a standard deviation of 91.87%. The average relative fair value is 69.86%, and the average relative overreaction of 30.14%.

[7] Some might suggest using the underpricing rate to measure IPO pricing efficiency. Given that we are focusing on IPO overreaction in this paper, we employ the Relative Overreaction (RO) to measure IPO pricing efficiency since RO not only aligns with our research goals but also provides readers with intuitive interpretation based on our research setting.

Penal B provides similar descriptive statistics for the STAR IPOs. The average initial return is 147.37% with a standard deviation of 150.89%. The median return is 111.23%. Even though the distribution is still skewed to the right the skewness level is less severe, compared to that for the registered ChiNext IPOs. The average monthly return is 132.18% with a standard deviation of 158.61%. The intramonth return is -15.19% with a standard deviation of 80.22%. It seems that the overreaction measured by intramonth return drops while the variation still remains high. The average relative fair value increases to 86.64%, which drops the average relative overreaction to 13.36%.

In the context of our study, the *RO* reflects the pricing error and thus is regarded as an important measure of IPO pricing efficiency. In comparing that between any two regulatory environments, we deem one that is more efficient than the other if the IPOs under that regulatory environment have, in the following particular order: 1) a statistically lower average of *RO*s; or 2) a statistically lower *RO* volatility if the averages of *RO*s are not statistically different. Following the above criteria, from Panel B in Table 2, we confirm that even though the mean *RO* for the 349 approved ChiNext IPOs is 13.87% that seems lower than 29.96% for the 393 registered ChiNext IPOs, their difference is not statistically significant (through a 2-tail t-test for equal means with a *p*-value = 0.60).[8] Furthermore, the variance of *RO* for the approved ChiNext IPOs reaches 3,176.21% that is significantly higher than 142.37% for the registered IPOs (through a 1-tail F-test for equal variances with a *p*-value = 0.00). These results expand Deng et al. (2023) in that the registered ChiNext IPOs are indeed priced more efficiently than the approved ones.

[8] We removed one outlier from both the approved and registered ChiNext IPO samples in the 2-tail t-test for the *RO* and *RFV* mean comparisons. Thus, the statistics for *RO* and *RFV* for both samples are slightly different from the corresponding measures for the full samples given in Table 1.

By comparing the registered ChiNext IPOs to STAR IPOs, and from Panel B in Table 3, we find that the mean of *RO*s for STAR IPOs is 13.36% that is significantly lower than that for ChiNext IPOs at 30.14% ($p$-value = 0.05). The variance of *RO* from ChiNext IPOs and STAR IPOs are not statistically different at the 5% level (but significant at the 10% level with the values of 142.62% and 165.50% along with a $p$-value of 0.06).[9] Using the criteria discussed above, we conclude that the registered STAR IPOs are priced more efficiently than the registered ChiNext IPOs.

## 4. Statistical Models and Empirical Results

### 4.1 Variable Screening and Multivariate Linear Regression Models

All the 57 explanatory variables for initial screening fall into four distinctive categories: 1) pre-listing demand, 2) post-listing demand, 3) market condition, and 4) pre-listing issuer value. To avoid possible multicollinearity, for any group of independent variables with high correlations ($|cor| \geq 0.400$), we run a regression to drop those that are not statistically significant at the 5% level ($p$-value $> 0.05$). If more than two highly correlated variables are statistically significant ($p$-value $\leq 0.05$) at the same time, we only keep the significant variable with the largest adjusted $R^2$ contribution. We repeat this pre-test procedure for all three return (dependent) variables for the ChiNext and STAR IPOs.

We then arrange IPOs according to their listing dates and use the following cross-sectional and multivariate OLS regression to analyze the impact, using the pre-screened significant variables on the initial return, monthly return and intramonth return for ChiNext and STAR IPOs, respectively:

$$IR_i/MR_i/IMR_i = \alpha_i + \sum_{j=1}^{n} \beta_j V_{i,j} + \epsilon_i \quad (6)$$

where $IR_i$ is the initial return, $MR_i$ is the monthly return, $IMR_i$ is the intramonth return, for IPO$_i$, $\alpha_i$ is the intercept, $\beta_j$ is the regression coefficient for explanatory variable $V_{i,j}$, and $\epsilon_i$ is an error term.

[9] We removed one outlier from both the registered ChiNext and registered STAR IPO samples in the F-test for the *RO* and *RFV* variance comparisons. Thus, the statistics for RO and RFV variances for both samples are slightly different from the corresponding measures for the full samples given in Table 1.

## 4.2 Empirical Results

Table 4 provides the comparisons among the approved and registered ChiNext IPOs, and registered STAR IPOs. We categorize all the significant variables for all three returns in different sample sets into four distinctive categories: 1) pre-listing demand, 2) post-listing demand, 3) market condition, and 4) pre-listing issuer value, and further discuss the roles and impacts they play on three returns in registration regime implementations.

### 4.2.1 ChiNext IPOs under Registration Regime

We follow the multivariate OLS regression on initial return, the return that pre-listing investors realize if they obtain the shares at the offer price and sell them at the listing day close price. From Panel A in Table 4 and for the registered ChiNext IPOs, we identify two significant variables. Both of them come from the pre-listing issuer value category: pre-issue P/E ratio (diluted) and the percentage of ownership from the largest Private Equity (PE) shareholders. The model produces an aggregated adjusted $R^2$ of 0.076.

The monthly return is what the pre-listing investors earn if they acquire the shares at the offer price and sell them at the closing price on the 21st trading day. It is a proxy of a short-term equilibrium return (Deng and Zhou, 2016) that reflects the issuer's fair value. From Panel B in Table 4, we find that the regression model for the monthly return produces a respectable adjusted $R^2$ of 0.611 and identifies four significant variables. Two of them are from the post-listing demand category and they are the range of returns and the listing day net capital inflow of medium-sized orders. The other two are from the pre-listing issuer value category, the pre-issue P/E ratio (diluted) and the ownership percentage of the largest PE shareholders.

The intramonth return is the return that post-listing traders earn if they acquire new shares at the closing price on the listing day and sell the shares at the closing price on the $21^{st}$ trading day. The intramonth return is different from the initial and monthly returns in that it reflects the two-way post-listing trading dynamics between the buyers (likely post-listing traders) and sellers (likely pre-listing investors). The buyers acquire shares to pursue future capital gains and therefore put on buy-support that drives up the price (evidenced by a positive standardized beta), while the sellers offload their shares to lock in the realized gains and therefore provide a sell-pressure to depress the price (evidenced by a negative standardized beta). As such, from the perspective of pre-listing investors, the intramonth return technically reflects the dissipation of overreaction rather than overreaction itself, while from the perspective of the post-listing traders, it is the potential return they could earn. As a result, we can treat the intramonth return as an outcome from a trading game between the pre-listing investors and post-listing traders with different expectations. A significant intramonth return indicates a significant overreaction embedded in the initial return. From Panel C in Table 4, we find that the regression model for the intramonth return yields an adjusted $R^2$ of 0.160 and identify three statistically significant variables. Two of them are from the post-listing demand category and they are the listing day P/E ratio and listing day net capital inflow from medium-sized orders. Another one is from the pre-listing issuer value catenary that is the pre-issue P/E ratio (diluted).[10]

### 4.2.2 STAR IPOs under Registration Regime

For STAR IPOs (Panel A in Table 4), we identify three significant variables for the initial return: percentage of ownership of the largest PE shareholders and pre-issue P/E ratio from the pre-listing issuer value category, and offline subscription that belongs to the pre-listing demand category. The model produces an aggregated adjusted $R^2$ of 0.038.

From Panel B in Table 4, we find that the regression model for the monthly return produces an adjusted $R^2$ of 0.725 and identifies four significant variables: range of returns, listing day net capital inflow from medium-sized orders, listing day net capital inflow from small-sized orders from the post-listing demand category, and the ranking of underwriting attorney firm from the pre-listing issuer value category.

From Panel C in Table 4, we find that the intramonth return is statistically significant, indicating that there is an overreaction embedded in the initial return. The regression model yields an adjusted $R^2$ of 0.065. There are three significant variables and all of them come from the post-listing category: 21-day turnover ratio, listing day net capital inflow from super-large-sized orders, and range of returns.

## 5. Impacts of Registration Regime Implementations on ChiNext and STAR IPOs

### 5.1 Initial Return

[10] Since the listing day P/E ratio refers to the P/E ratio on the listing day and thus it belongs to the post-listing demand category. The correlation coefficient matrix (not reported in this paper to save space) shows that the listing day P/E ratio and pre-issue P/E ratio (diluted) for the registered ChiNext IPOs in the entire sample period is -0.358.

The initial return of the registered ChiNext IPOs is entirely determined by the variables in the pre-listing issuer value category. With an updated dataset our results further validate Deng et al. (2023). For the registered STAR IPOs, the initial return is determined by both pre-listing issuer value and pre-listing demand. By comparing the registered ChiNext IPOs and STAR IPOs, we find that the difference is that while the pre-listing issuer value is significant for both, the pre-listing demand is only significant for the STAR IPOs (Panel A in Table 4). This can be explained from two angles: 1) listing rules for issuers, and 2) participation requirements for investors.

First, the STAR issuers face more specific and quantifiable listing rules than their ChiNext counterparts, particularly for those with lower expected market cap (refer to Panel C in Appendix 1). For example, an issuer that aspires to list on ChiNext needs to meet one of three "loosely defined" criteria, while if it wishes to get listed on STAR it must satisfy one of five very specifically designed conditions, and more specific listing rules provide clearer and more appliable guidelines for the investors to evaluate the issuers. As a result, the more specific requirements for the STAR IPO issuers help mitigate the uncertainty faced by the investors and make the pricing more efficient.

Second, while the STAR investors are still not as experienced in examining the issuers' prospectuses as the regulators, they are in general more "sophisticated" than their ChiNext counterparts (refer to Panel C in Appendix 1). Specifically, the ChiNext investors are only required to have 100,000 Yuan on average of financial assets 20 days prior to trading, while the same number for the STAR investors is 500,000 Yuan.[11] As such, the STAR investors are of higher net worth in general and thus more sophisticated; they purposefully acquire valuable information from their peer group through the pre-listing demand variables. They are particularly interested in the average offline subscription, which reflects the collective demand from the more sophisticated investor group. Therefore, it is reasonable to assume that more sophisticated investors disseminating easier-to-digest information have greater potential to reduce pricing errors. That explains why the relative overreaction for the STAR IPOs is significantly lower than that for the ChiNext IPOs (13.36% vs. 30.14%), which provides a strong support to this argument.

[11] Yuan is the unit of Chinese currency. The exchange rate between the US Dollar and China Yuan is around $1 for 7.25 Yuan in early July 2023.

**5.2 Monthly Return (Fair Value)**

Deng and Zhou (2016) demonstrate that the monthly return represents a short-term equilibrium return, thus it is the objective measure of the issuers' fair value by the market. For the registered ChiNext IPOs, the monthly return is essentially explained by post-listing demand (adjusted $R^2$ of 0.571), with a minor adjustment from pre-listing issuer value (adjusted $R^2$ of 0.040) (Panel B in Table 4).

For the registered STAR IPOs, the monthly return is largely determined by post-listing demand (adjusted $R^2$ of 0.723) with a minor adjustment from pre-listing issuer value (adjusted $R^2$ of 0.002) (Panel B in Table 4). It is interesting that the significant variables for the fair value for both registered ChiNext and STAR IPOs are from the same variable categories. This indicates that the market demand provides the most accurate assessment on the issuers' fair value for the registered IPOs for both ChiNext and STAR.

**5.3 Intramonth Return (Overreaction)**

Deng and Zhou (2016) establish that the intramonth return is a proxy of overreaction in the initial return. More precisely, in the context of our model, it reflects the dissipation of overreaction during the first month of trading. For the registered ChiNext IPOs the dissipation of overreaction is decided by post-listing demand (adjusted $R^2$ of 0.098) and then pre-listing issuer value (adjusted $R^2$ of 0.061) (Panel C in Table 4). The relative overreaction (RO) is high at 30.14% (Panel A in Table 1), relative to the initial return, a strong indication that the investors are less capable of discovering the issuers' fair value, hence there is sizeable pricing error that manifests as the overreaction. The market plays a key role in pricing discovery and error correction. A closer look at the signs and numerical values of the significant variables suggests that the sell-pressure from the sellers (negative standard betas) dominates the buy-support from the buyers (positive standardized betas). This provides a partial answer to why the relative level of overreaction is higher for the registered ChiNext IPOs compared to STAR IPOs.

For registered STAR IPOs, the dissipation of overreaction is decided entirely by post-listing demand (adjusted $R^2$ of 0.065) (Panel C in Table 4). There are two noticeable differences in comparing to the registered ChiNext IPOs: the relative level of overreaction for the registered STAR IPOs is significantly lower (13.36% vs. 30.14%) (Panel B in Table 3) and the dissipation of overreaction for the registered STAR IPOs is no longer affected by the pre-listing issuer value. And again, these results can be explained from the angles of listing rules for the issuers and participation requirements for the investors.

The registered STAR issuers have a lower level of perceived risk because of more specific listing requirements than the registered ChiNext IPOs, and the registered STAR investors are in general more sophisticated than their ChiNext colleagues. As such, the dissemination of information embedded in the pre-listing issuer value is more efficient and of a higher quality, which in turn is fully consumed in establishing the initial return and fair value, thus is not wasted in creating overreaction. Therefore, the registered STAR IPOs enjoy much lower relative level of overreaction. The dissipation of overreaction is entirely determined by the secondary market trading dynamics reflected by the post-listing demand. The signs and numerical values of the significant variables suggest that the buy-support side (positive standardized betas) dominates the dynamics of secondary trading. There are actually no significant variables to represent the sell-pressure side (negative standard betas). This result provides another partial answer to why the relative overreaction is lower for the registered STAR IPOs.

**5.4 IPO Pricing Efficiency**

Deng et al. (2023) thoroughly analyze the impacts of ChiNext IPO regulation regime change from an approval one to a registration one. They find that the regime change has caused a fundamental paradigm shift of ChiNext IPO pricing from demand-driven to value-driven, and that the impacts of regulation reforms on IPO returns can largely be explained by three factors: 1) governing regulation regime, 2) listing day trading restrictions, and 3) issuer profile. With an updated dataset, we extend their findings in that the variance of relative overreaction for the registered ChiNext IPOs is significantly lower than that of the approved IPOs (142.37% vs. 3,176.21%) (Panel B in Table 2). Furthermore, we also find that the relative overreaction is much lower for the registered STAR IPOs than that for the registered ChiNext IPOs (13.36% vs. 30.14%) (Panel B in Table 3), and that the impacts of the two registration regime implementations on IPO returns are closely related to two factors: 1) the listing rules for issuers, and 2) the participation requirements for investors.

The STAR board is the first board in China that adopts a full registration process at the very beginning of its existence. It caters to budding issuers that lean towards "hard technology," particularly those with strong R&D potential, even in their earlier stage of development and thus deemed as "riskier" than their counterparts listed on the "more established" ChiNext board. Therefore, while the regulators still have the mandate to make the IPO market more accessible to early-stage technology companies in a timely fashion, they need to balance their efforts with strengthened investor protection measures. They achieve the balanced goal by being more specific on listing rules for the issuers (SHSE Bulletin #[2020]101 on December 31, 2020) and raising the bar of participation for the investors (SHSE Bulletin #[2019]23 on Mar 1, 2019) (refer to Panel C in Appendix 1).

As Deng et al. (2023) have established, the key deficiency for the registered ChiNext IPOs is that less sophisticated investors are left to disseminate the information from prospectuses with a lower level of regulatory scrutiny, which creates a higher level of overreaction that results in less efficient pricing. The registration regime for the STAR IPOs seems to provide a better alternative primarily because it corrects the deficiency in the ChiNext IPOs in a very targeted way by demanding 1) more specific listing rules for the issuers, and 2) more stringent participation requirements for the investors.

A question on IPO pricing efficiency remains to be answered: which regulation regime produces more efficient IPO pricing? Deng et al. (2023) document that the registration regime is more efficient than the approval regime in pricing ChiNext IPOs, because the value-driven pricing process under the registration regime is more market-oriented in that the fair value and dissipation of overreaction are almost entirely determined by the secondary market trading dynamics. Our results not only support Deng et al. (2023) in that the registration regime produces more efficient IPO pricing than the approval regime, but also further look into the differences in listing and participations requirements of the issuers and investors.

Our results suggest that the registration regime itself can be carefully designed, that an implementation with more specific listing requirements for the issuers and more stringent participation requirements for the investors leads to a lower relative overreaction and therefore a higher efficiency in IPO pricing. In that regard, our conclusion is that among the three IPO regulation regimes for the two entrepreneurial boards (approved ChiNext, registered ChiNext, and registered STAR), the registration regime governing the STAR IPOs offers the highest IPO pricing efficiency. Synthesizing our results with those reported in Deng et al. (2023), we propose that the most pricing efficient IPO regulation regime has the following four characteristics: 1) registration system, 2) no hard return caps and no trading curbs that cap the initial return; 3) more specific listing rules for the issuers, and 4) more stringent participation requirements for the investors.

It is worth mentioning that in order to reduce a "free-ride" risk, after the establishment of registration regime for both ChiNext and STAR, underwriters have been required to discard the highest-priced bids from institutional investors when determining a new issue's offer price. Before September 18, 2021, the practice was to exclude no less than the highest 10% of the bids from institutional investors. On September 18, 2021, with the release of the SZSE Bulletin #[2021]919 and SHSE Bulletins #[2021]76 and #[2021]77 (the 2021 reform), underwriters are required to reject no more than the highest 3% of the institutional bids for both boards. In addition, the upper bound of the offer price range cannot be higher than the lower bound by 20%. The net effect is that, for both ChiNext and STAR, after the 2021 reform, offer price is higher, resulting in lower initial return, monthly return and intramonth return, than those before the reform. On the other hand, we have observed that the relative fair value and relative overreaction are much higher after the 2021 reform than the corresponding measures prior to the reform. The impacts of the 2021 reform on IPO pricing efficiency, and in general, the impacts of IPO pricing mechanism design on IPO pricing efficiency, would be a highly desirable topic for a follow-up study.

## 6. Conclusions

This study examines the variations of the two registration regime implementations and their impacts on initial return of IPOs listed on the ChiNext and STAR boards. We analyze the two fundamental components in IPO initial return: an issuer's fair value represented by the IPO's monthly return known as a short-term equilibrium return, and an overreaction from the investors represented by its intramonth return, which is the difference between the monthly return and initial return. We compare the levels of initial return, fair value, and overreaction for ChiNext IPOs and STAR IPOs, as well as the determinants that drive them to understand the underlying factors causing the differences. In addition, we seek to expand Deng et al. (2023) by providing updated empirical evidence on ChiNext and STAR IPOs' pricing efficiency.

We find that all the significant variables that affect the initial return, monthly return, and intramonth return for ChiNext and STAR IPOs fall into four categories: 1) pre-listing demand, 2) post-listing demand, 3) market condition, and 4) pre-listing issuer value. We observe high similarity among the variable categories for each of the returns between the IPOs from two different boards under the registration regime; moreover, the STAR IPOs exhibit lower relative level of overreaction. Our findings indicate that the differences between the levels and determinants of initial return, fair value, and overreaction for the ChiNext and STAR IPOs can largely be explained by the two implementation variations: 1) listing rules for issuers, and 2) participation requirements for investors.

We document that the level of relative overreaction for registered STAR IPOs is lower than that for registered ChiNext IPOs, proving that the registration regime implementation governing the STAR IPOs is more efficient in IPO pricing. We also find that the variance of relative overreaction for the registered ChiNext IPOs is lower than that for the approved ones, which provides direct empirical evidence to support that the registration regime yields more efficiently priced IPOs than the approval regime. We conclude that the most efficient regulation regime in IPO pricing has the following four characteristics: 1) registration system, 2) no hard return caps and nor trading curbs that will cap the initial return, 3) more specific listing rules for the issuers, and 4) more stringent participation requirements for the investors.

As the CSRC continues fine-tuning the IPO process to further increase pricing efficiency and to reduce potential risk (volatility) in order to protect investors, our findings provide insight to help policymakers achieve their goals. Our study also casts new lights for future research to enhance the transparency and efficiency of the Chinese IPO markets.

## References

**The following papers are published on English language journals:**

**The following papers are published on Chinese language journals:**

**Figure 1 – Distributions of Initial Return, Monthly Return, and Intramonth Return for ChiNext IPOs**

This figure illustrates the distributions (boxplots and histograms) of initial return (IR), monthly return (MR), and intramonth return (IMR) for 393 registered ChiNext IPOs from August 24, 2020 to November 11, 2022 with one extreme outlier removed. The returns are measured in percentages.

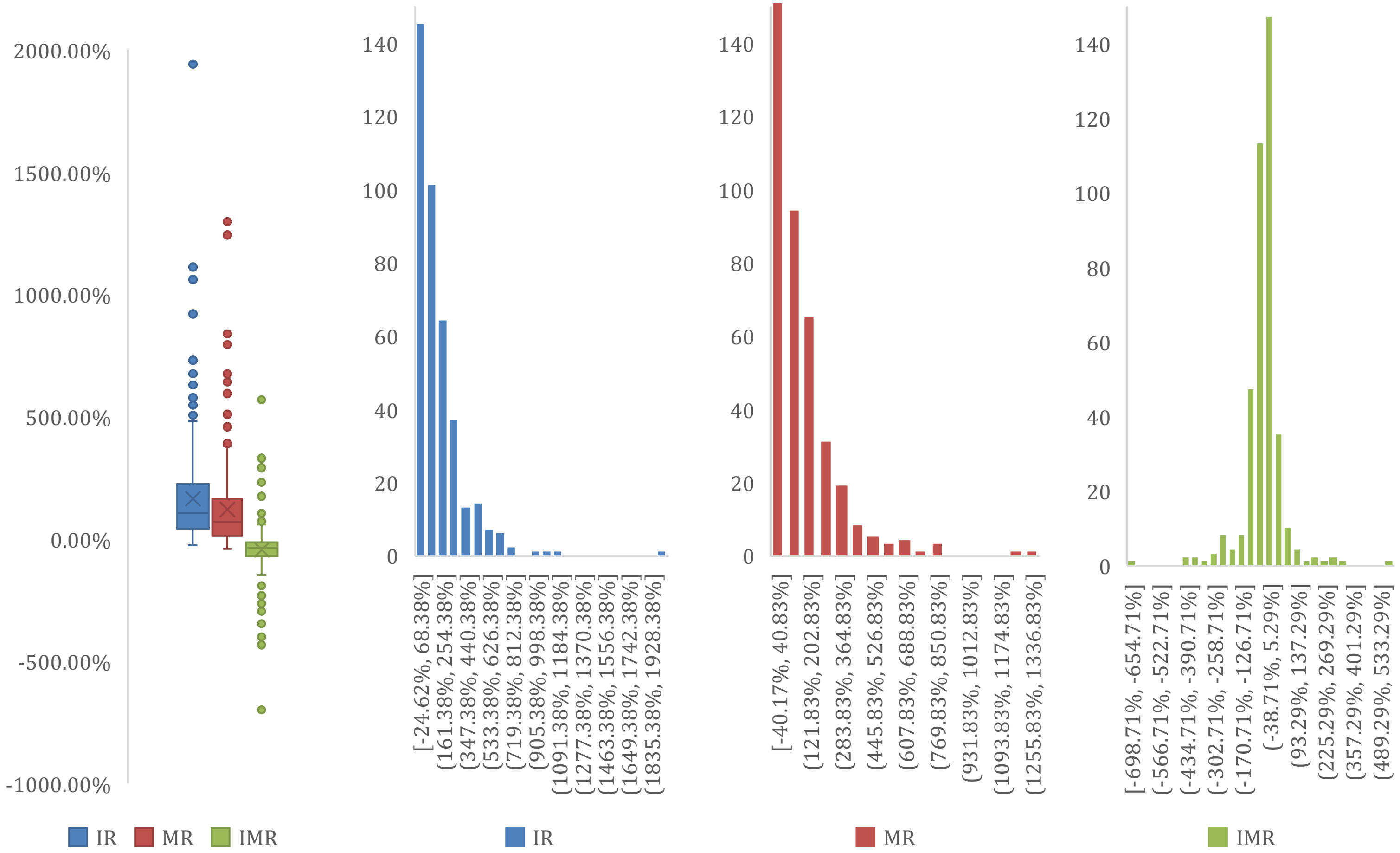

**Figure 2 – Distributions of Relative Fair Value and Relative Overreaction for ChiNext IPOs**

This figure illustrates the distributions (boxplots and histograms) of relative fair value (RFV) and relative overreaction (RO) for 393 registered ChiNext IPOs from August 24, 2020 to November 11, 2022 with one extreme outlier removed. Both RFV and RO are measured in percentages.

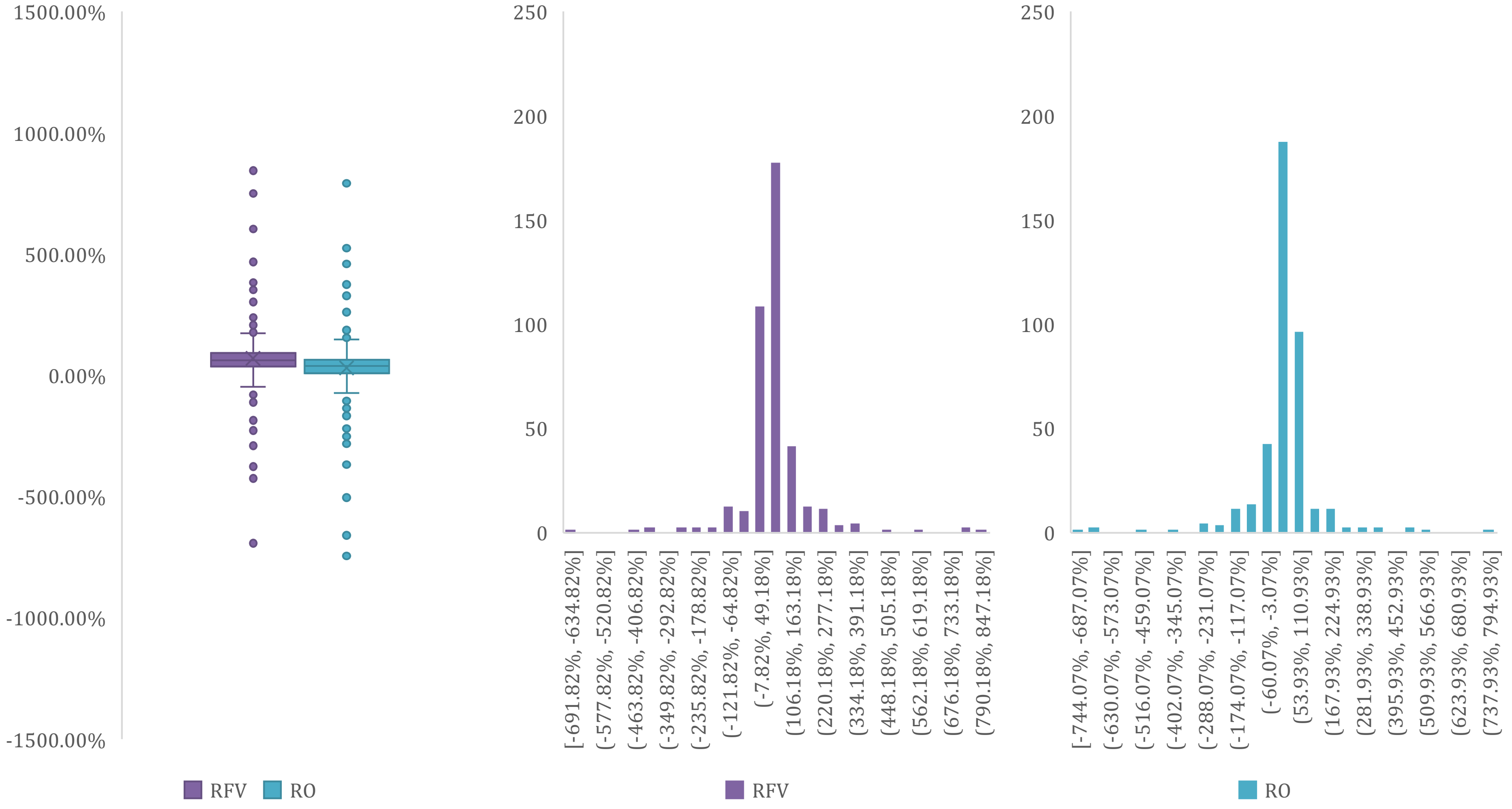

Note 1) RFV – Relative Fair Value; RO – Relative Overreaction

**Figure 3 – Distributions of Initial Return, Monthly Return, and Intramonth Return for STAR IPOs**

This figure illustrates the distributions (boxplots and histograms) of initial return (IR), monthly return (MR) and intramonth remove (IMR) for 451 registered STAR IPOs from July 22, 2019 to August 18, 2022 (with one extreme outlier removed). The returns are measured in percentages.

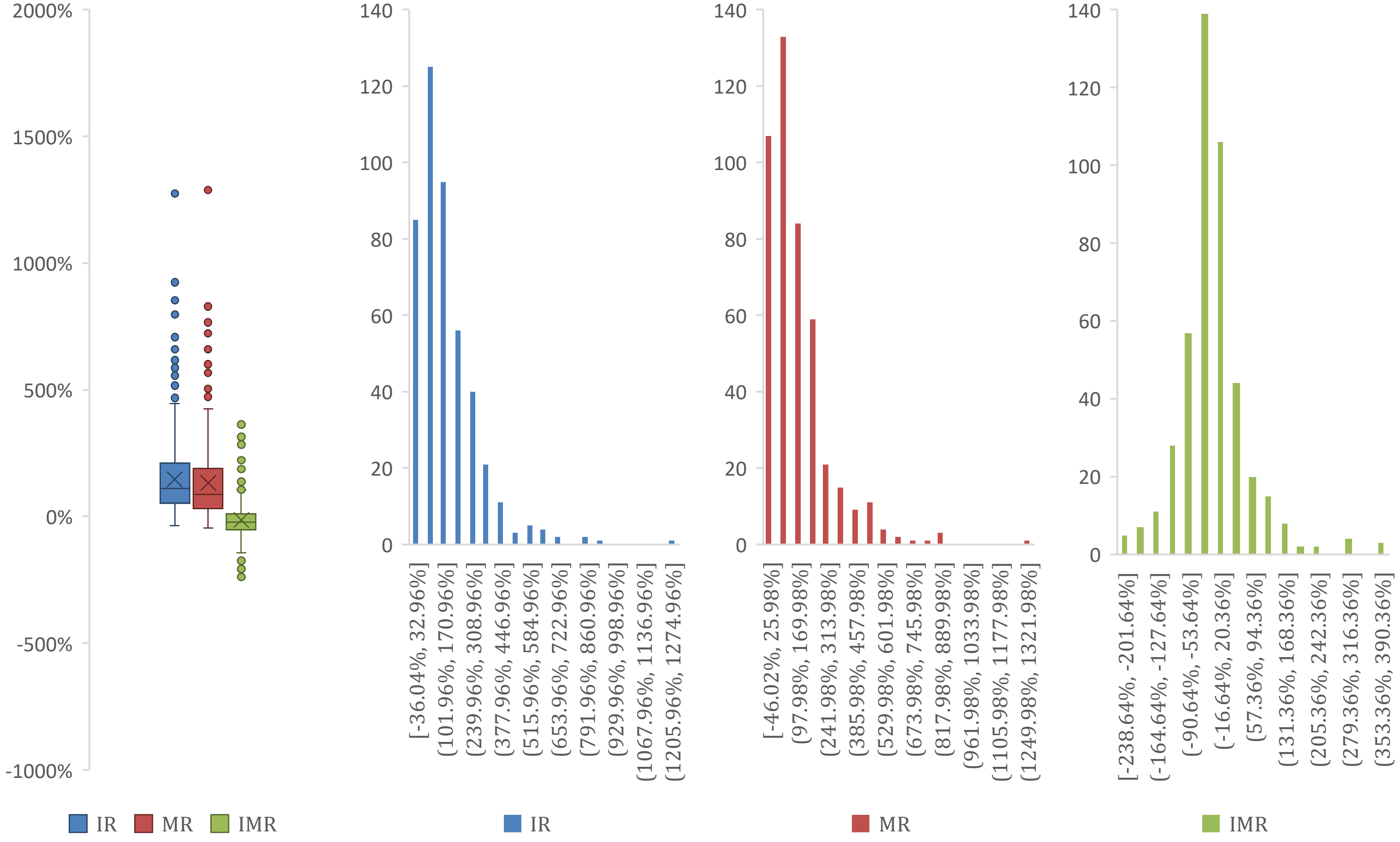

Note 1) IR – Initial Return; MR – Monthly Return; IMR – Intramonthly Return

## Figure 4 - Distributions of Relative Fair Value and Relative Overreaction for STAR IPOs

This figure illustrates the distributions (boxplots and histograms) of relative fair value (RFV) and relative overreaction (RO) for 451 registered STAR IPOs from July 22, 2019 to August 18, 2022 with one extreme outlier removed. The RFV and RO are measured in percentages.

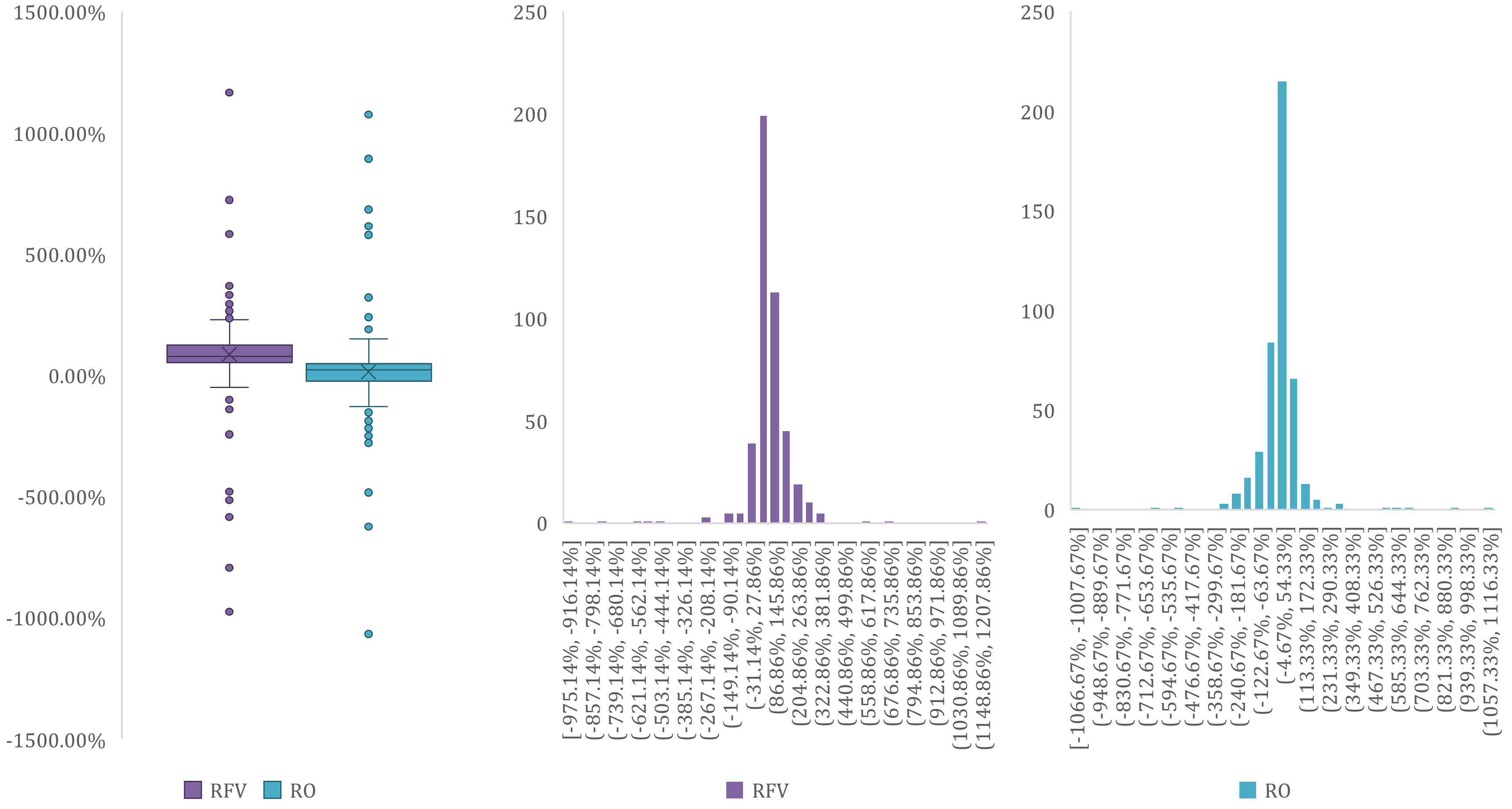

Note 1) RFV – Relative Fair Value; RO – Relative Overreaction

## Table 1 - Descriptive Statistics of Initial, Monthly, and Intramonth Returns

We report the descriptive statistics of initial return (IR), monthly return (MR), intramonth return (IMR), relative fair value (RFV) and relative overreaction (RO) for 393 registered ChiNext IPOs (with one extreme outlier removed) from August 24, 2020 to November 11, 2022 in Panel A and for 451 registered STAR IPOs (with one extreme outlier removed) from July 22, 2019 to August 18, 2022 in Panel B.

**Panel A - Registered ChiNext IPO Descriptive Statistics**

| ChiNext | IR | MR | IMR | RFV | RO |
|---|---|---|---|---|---|
| Mean | 164.79% | 121.06% | -43.73% | 69.86% | 30.14% |
| Median | 104.91% | 70.63% | -36.11% | 61.99% | 38.01% |
| Standard Deviation | 194.41% | 169.24% | 91.87% | 119.32% | 119.32% |
| Sample Variance | 377.94% | 286.42% | 84.39% | 142.37% | 142.37% |
| Kurtosis | 20.18 | 12.63 | 14.55 | 16.07 | 16.07 |
| Skewness | 3.29 | 2.88 | -0.47 | 0.92 | -0.92 |
| Count | 393 | 393 | 393 | 393 | 393 |

**Panel B - Registered STAR IPO Descriptive Statistics**

| STAR | IR | MR | IMR | RFV | RO |
|---|---|---|---|---|---|
| Mean | 147.37% | 132.18% | -15.19% | 86.64% | 13.36% |
| Median | 111.23% | 86.83% | -21.76% | 78.02% | 21.98% |
| Standard Deviation | 150.89% | 158.61% | 80.22% | 128.52% | 128.52% |
| Sample Variance | 227.69% | 251.56% | 64.36% | 165.17% | 165.17% |
| Kurtosis | 10.12 | 9.21 | 4.95 | 29.89 | 29.89 |
| Skewness | 2.39 | 2.40 | 1.27 | -0.89 | 0.89 |
| Count | 451 | 451 | 451 | 451 | 451 |

## Table 2 - Comparisons of Relative Fair Value and Relative Overreaction: Approved ChiNext verse Registered ChiNext IPOs

We report the results for the sample mean tests (assuming different variances) and for the sample variance tests for the relative fair value (*RFV*) and relative overreaction (*RO*) to compare the approved ChiNext IPOs with 349 observations from October 30, 2009 to December 31, 2012 (Approved with one extreme outlier removed) and the registered ChiNext IPOs with 393 observations from August 24, 2020 to November 11, 2022 (Registered with two outliers removed). Panel A reports the test results for *RFV* and Panel B reports the results for *RO*. The tests were conducted with Excel, and ***, **, and * indicate significant at the 1%, 5%, and 10% levels, respectively.

**Panel A - Relative Fair Value (RFV) Comparisons**

| ChiNext | t-Test: Two-Sample | | ChiNext | F-Test: Two-Sample | |
|---|---|---|---|---|---|
| RFV | Approved | Registered | RFV | Approved | Registered |
| Mean | 86.13% | 70.04% | Variance | 3176.21% | 142.37% |
| Observations | 348 (1) | 392 (2) | Observations | 349 | 393 |
| df | 375 | | df | 348 | 392 |
| t Stat | 0.52 | | F | 22.31 | |
| *p* (T<=t) two-tail | 0.60 | | *p* (F<=f) one-tail | 0.00*** | |
| t Critical two-tail | 1.97 | | F Critical one-tail | 1.19 | |

**Panel B - Relative Overreaction (RO) Comparisons**

| ChiNext | t-Test: Two-Sample | | ChiNext | F-Test: Two-Sample | |
|---|---|---|---|---|---|
| RO | Approved | Registered | RO | Approved | Registered |
| Mean | 13.87% | 29.96% | Variance | 3176.21% | 142.37% |
| Observations | 348 (1) | 392 (2) | Observations | 349 | 393 |
| df | 375 | | df | 348 | 392 |
| t Stat | -0.52 | | F | 22.31 | |
| *p* (T<=t) two-tail | 0.60 | | *p* (F<=f) one-tail | 0.00*** | |
| t Critical two-tail | 1.97 | | F Critical one-tail | 1.19 | |

Notes (1) This t-test was conducted with Excel, which removed one additional outlier deemed to be extreme from the Approved ChiNext IPO sample set with 349 samples.
Notes (2) This t-test was conducted with Excel, which removed one additional outlier deemed to be extreme from the registered ChiNext IPO sample set with 393 samples.

## Table 3 - Comparisons of Relative Fair Value and Relative Overreaction: Registered ChiNext IPOs verse Registered STAR IPOs

We report the test results for the sample mean tests (assuming different variances) and for the sample variance tests for the relative fair value (RFV) and relative overreaction (RO) to compare the registered ChiNext IPOs with 393 observations from August 24, 2020 to November 11, 2022 (with one extreme outlier removed) and STAR IPOs with 451 observations from July 22, 2020 to August 18, 2022 (with one extreme outlier removed). Panel A reports the test results for RFV and Panel B reports the results for RO. The tests were conducted with Excel and ***, **, and * indicate significant at the 1%, 5%, and 10% levels, respectively.

**Panel A - Relative Fair Value (RFV) Comparisons**

| Registered | t-Test: Two-Sample | | Registered | F-Test: Two-Sample | |
|---|---|---|---|---|---|
| RFV | ChiNext | STAR | RFV | ChiNext | STAR |
| Mean | 69.86% | 86.64% | Variance | 142.62% | 165.50% |
| Observations | 393 | 451 | Observations | 392 (1) | 450 (2) |
| df | 839 | | df | 391 | 449 |
| t Stat | 1.97 | | F | 1.16 | |
| *p* (T<=t) two-tail | 0.05** | | *p* (F<=f) one-tail | 0.06* | |
| t Critical two-tail | 1.96 | | F Critical one-tail | 1.18 | |

**Panel B - Relative Overreaction (RO) Comparisons**

| Registered | t-Test: Two-Sample | | Registered | F-Test: Two-Sample | |
|---|---|---|---|---|---|
| RO | ChiNext | STAR | RO | ChiNext | STAR |
| Mean | 30.14% | 13.36% | Variance | 142.62% | 165.50% |
| Observations | 393 | 451 | Observations | 392 (1) | 450 (2) |
| df | 839 | | df | 391 | 449 |
| t Stat | -1.97 | | F | 1.16 | |
| *p* (T<=t) two-tail | 0.05** | | *p* (F<=f) one-tail | 0.06* | |
| t Critical two-tail | 1.96 | | F Critical one-tail | 1.18 | |

Notes (1) This F-test was conducted with Excel, which removed one additional outlier deemed to be extreme from the registered ChiNext IPO sample set with 393 samples.
Notes (2) This F-test was conducted with Excel, which removed one additional outlier deemed to be extreme from the registered STAR IPO sample set with 451 samples.

## Table 4 - Comparisons of Determinant Categories of Returns: Registered ChiNext vs. Registered STAR IPOs

We provide the significant variables and their statistics from regression (6) for the initial return (Panel A), monthly return (panel B), and intramonth return (Panel C) for 393 ChiNext IPOs from August 24, 2020 to November 11, 2022 and for 451 STAR IPOs from July 22, 2019 to August 18, 2022. We compare the results in this table since both types of IPOs are under the same registered regime.

**Panel A – Initial Return (IR): Comparisons of Determinant Categories in Three Sample Sets**

| Initial Return: Mean: 164.79%, SD: 194.41% Model Fit and Significant Variables | ChiNext IPOs under Registration | | |
|---|---|---|---|
| | Standardized Beta | Adjust $R^2$ Change | Group Adjust $R^2$ |
| Pre-listing Demand | | | |
| Post-listing Demand | | | |
| Market Condition | | | |
| Pre-listing Issuer Value | | | |
| Pre-issue P/E ratio (diluted) | -0.263 | 0.065 | 0.076 |
| Ownership of largest PE shareholder (%) | 0.121 | 0.011 | |
| Adjusted $R^2$ | 0.076 | | |

| Initial Return: Mean: 147.41%, SD: 151.06% Model Fit and Significant Variables | STAR IPOs under Registration | | |
|---|---|---|---|
| | Standardized Beta | Adjusted $R^2$ Change | Group Adjust $R^2$ |
| Pre-listing Demand | | | |
| Offline subscription ratio (average) | -0.130 | 0.012 | 0.012 |
| Post-listing Demand | | | |
| Market Condition | | | |
| Pre-listing Issuer Value | | | |
| Ownership of largest PE shareholder (%) | 0.163 | 0.014 | 0.026 |
| Pre-issue P/E ratio (diluted) | -0.131 | 0.012 | |
| Adjusted $R^2$ | 0.038 | | |

# Table 4 (con'd) - Comparisons of Determinant Categories of Returns: Registered ChiNext vs. Registered STAR IPOs

We provide the significant variables and their statistics from regression (6) for the initial return (Panel A), monthly return (panel B), and intramonth return (Panel C) for 393 ChiNext IPOs from August 24, 2020 to November 11, 2022 and for 451 STAR IPOs from July 22, 2019 to August 18, 2022. We compare the results in this table since both types of IPOs are under the same registered regime.

**Panel B - Monthly Return (MR): Comparisons of Determinant Categories in Three Sample Sets**

| Monthly Return<br>Mean: 121.06%, SD: 169.24%<br>Model Fit and Significant Variables | ChiNext IPOs under Registration<br>Standardized Beta | Adjust $R^2$ Change | Group Adjust $R^2$ |
|---|---|---|---|
| Pre-listing Demand | | | |
| Post-listing Demand | | | |
| Range of returns (%) | 0.713 | 0.524 | 0.571 |
| Listing day net capital inflow of medium-sized orders | -0.238 | 0.046 | |
| Market Condition | | | |
| Pre-listing Issuer Value | | | |
| Pre-issue P/E ratio (diluted) | -0.182 | 0.032 | 0.040 |
| Ownership of largest PE shareholder (%) | -0.098 | 0.008 | |
| Adjusted $R^2$ | 0.611 | | |

| Monthly Return<br>Mean: 132.11%, SD: 158.78%<br>Model Fit and Significant Variables | STAR IPOs under Registration<br>Standardized Beta | Adjusted $R^2$ Change | Group Adjust $R^2$ |
|---|---|---|---|
| Pre-listing Demand | | | |
| Post-listing Demand | | | |
| Range of returns (%) | 0.813 | 0.688 | 0.723 |
| Listing day net capital inflow of medium-sized orders | -0.171 | 0.023 | |
| Listing day net capital inflow of small-sized orders | -0.117 | 0.012 | |
| Market Condition | | | |
| Pre-listing Issuer Value | | | |
| Ranking of underwriting attorney firm | 0.051 | 0.002 | 0.002 |
| Adjusted $R^2$ | 0.725 | | |

## Table 4 (con'd) - Comparisons of Determinant Categories of Returns: Registered ChiNext vs. Registered STAR IPOs

We provide the significant variables and their statistics from regression (6) for the initial return (Panel A), monthly return (panel B), and intramonth return (Panel C) for 393 ChiNext IPOs from August 24, 2020 to November 11, 2022 and for 451 STAR IPOs from July 22, 2019 to August 18, 2022. We compare the results in this table since both types of IPOs are under the same registered regime.

**Panel C - Intramonth Return (IMR): Comparisons of Determinant Categories in Three Sample Sets**

| Intramonthly Return<br>Mean: -43.73%, SD: 91.87%<br>Model Fit and Significant Variables | ChiNext IPOs under Registration<br>Standardized Beta | <br>Adjust $R^2$ Change | <br>Group Adjust $R^2$ |
|---|---|---|---|
| Pre-listing Demand | | | |
| Post-listing Demand | | | |
| Listing day P/E ratio | -0.406 | 0.063 | 0.098 |
| Listing day net capital inflow of medium-sized orders | -0.199 | 0.035 | |
| Market Condition | | | |
| Pre-listing Issuer Value | | | |
| Pre-issue P/E ratio (diluted) | 0.267 | 0.061 | 0.061 |
| Adjusted $R^2$ | 0.160 | | |

| Intramonthly Return<br>Mean: -15.30%, SD: 80.28%<br>Model Fit and Significant Variables | STAR IPOs under Registration<br>Standardized Beta | <br>Adjusted $R^2$ Change | <br>Group Adjust $R^2$ |
|---|---|---|---|
| Pre-listing Demand | | | |
| Post-listing Demand | | | |
| 21-day turnover ratio | 0.191 | 0.044 | |
| Listing day net capital inflow of super-large-sized orders | 0.109 | 0.014 | 0.065 |
| Range of returns | 0.107 | 0.007 | |
| Market Condition | | | |
| Pre-listing Issuer Value | | | |
| Adjusted $R^2$ | 0.065 | | |

## Appendix 1 - Comparisons of Regulation Regimes and Their Impacts on ChiNext IPOs

We compare the categories of the significant variables for the initial return, monthly return, and intramonth return across three sample periods (ChiNext IPOs under approval regime, ChiNext IPOs under registration regime, and STAR IPOs under registration regime) along with the rules and regulations (restrictions) imposed on the IPO firms, initial investors (pre-IPO), regular traders (post-IPO) in different sample period in Panels A-C.

**Panel A - Regulation Regime Comparisons**

| Sample Set<br>Regulation Regime<br>Time Period | ChiNext under Approval | ChiNext under Registration | STAR under Registration |
|---|---|---|---|
| | Prior to 2013 Market Reform | Post 2020 ChiNext Reform | Post 2019 STAR Reform |
| | October 30, 2009 – December 31, 2012 | August 24, 2020 – November 11, 2022 | July 22, 2019 – August 18, 2022 |
| Regulation regime | Approval | Registration | Registration |
| Regulator mandate | Foster capital market growth | Satisfy company capital needs | Satisfy company capital needs |
| Regulator emphasis | Investor protection | Compliance | Compliance |
| IPO pricing mechanism | Demand-driven | Value-driven | Value-driven |
| Overreaction level (relative) | 13.87% | 30.14% | 13.36% |
| Listing day trading restrictions (See Panel B) | Intraday trading curbs | No hard return caps or trading curbs that cap return | No hard return caps or trading curbs that cap return |
| Issuer listing rules (see Panel C) | Not specific | Less specific | Most specific |
| Investor participation Requirements (see Panel C) | None | 100,000 RMB | 500,000 RMB |
| Regulation post-listing | Weak | Strong | Strong |
| Self-regulation/Self-discipline | Weak | Strong | Strong |
| Delisting rules | None | Yes | Yes |
| IPO Pricing Efficiency | Low | Medium | High |

**Panel B - Listing Day Trading Restrictions Comparisons**

| Sample Set<br>Regulation Regime<br>Time Period | ChiNext Approval<br>Prior to 2013 Market Reform<br>Oct 30, 2009 – Dec 31, 2012 | ChiNext Registration<br>Post 2020 ChiNext Reform<br>Aug 24, 2020 – Nov 11, 2022 | STAR Registration<br>Post 2019 STAR Reform<br>Jul 22, 2019 – Aug18, 2022 |
|---|---|---|---|
| Listing day return caps | 1) no listing day return cap;<br>2) 10% return cap on subsequent trading days. | 1) no return cap for first five trading days including listing day;<br>2) 20% return cap on subsequent trading days. | 1) no return cap for first five trading days including listing day;<br>2) 20% return cap on subsequent trading days. |
| Listing day trading curbs | 1) first price suspension if intraday return at or above 10% for 1 hour;<br>2) second price suspension if intraday return at or above 20% till 14:57;<br>3) if turnover ratio at or above 50%, suspension for 1 hour;<br>4) if a 1-hour suspension goes beyond 14:57, trading resume at 14:57 for the closing call auction. | 1) first suspension if intraday return is above 30%, second suspension if intraday return is above 60%;<br>2) intraday suspension lasts 10 minutes;<br>3) if a 10-minute suspension goes beyond 14:57, trading resume at 14:57 for the closing call auction. | not specified |
| Subsequent trading day abnormal trades | 1) abnormal trade: accumulative return deviation (= accumulative return - accumulative ChiNext Index return) reaches 20% for three consecutive trading days;<br>2) ratio between [average 3-day turnover ratio] and previous [5-day turnover ratio] reaches 30, and turnover ratio is no less than 20% per day for the 3 days. | 1) abnormal trade: accumulative return deviation (= accumulative return - accumulative ChiNext Index return) reaches 30% for three consecutive trading days;<br>2) severe abnormal trade: item 1) happens 3 times in 10 consecutive trading days.<br>3) severe abnormal trade: accumulative return deviation reaches +100% (-50%) in 10 consecutive trading days;<br>4) severe abnormal trade: accumulative return deviation is reaches +200% (-70%) in 30 consecutive trading days. | 1) abnormal trade: accumulative return deviation (= accumulative return - accumulative STAR Index return) reaches 30% for three consecutive trading days. |
| After-market trade at closing price | no | 1) single order at closing price after market close no more than 1,000,000 shares. | 1) single order at closing price after market close no more than 1,000,000 shares. |
| Number of shares limitation per order | 1) single order through auctions (call and continuous) no more than 1,000,000 shares. | 1) single limit order through auctions (call and continuous) no more than 300,000 shares;<br>2) single market order through auctions (call and continuous) no more than 150,000 shares. | 1) single limit order through auctions (call and continuous) no more than 100,000 shares;<br>2) single market order through auctions (call and continuous) no more than 50,000 shares. |

**Panel C - Issuer and Investor Profile Comparisons**

| Sample Set<br>Regulation Regime<br>Time Period | ChiNext Approval<br>Prior to 2013 Market Reform<br>Oct 30, 2009 – Dec 31, 2012 | ChiNext Registration<br>Post 2020 ChiNext Reform<br>Aug 24, 2020 – Nov 11, 2022 | STAR Registration<br>Post 2019 STAR Reform<br>Jul 22, 2019 – Aug 18, 2022 |
|---|---|---|---|
| Issuer listing rules | Meet one of following three conditions:<br>1. positive net profit for last 2 years with total profit no less than 10M RMB, and in growth;<br>2. positive net profit for last 1 year with total net profit no less than 5M RMB and revenue no less 50M RMB, and annualized revenue growth rate no lower than 30% per year;<br>3. total net asset of last accounting cycle no less than 20M RMB. | Meet one of following three conditions:<br>1. positive net profit for last 2 years with total net profit no less than 50M RMB;<br>2. expected market cap no lower than 1B RMB, positive net profit for last 1 year with revenue no lower than 100M RMB;<br>3. expected market cap no lower than 5B RMB, revenue no lower than 300M RMB for last 1 year. | Meet one of following three conditions:<br>1. expected market cap no lower than 1B RMB, positive net profit for last 2 years with accumulated profit no lower than 50M RMB; or expected market cap no lower than 1B RMB; positive net profit for last 1 year with revenue no lower than 100M RMB;<br>2. expected market cap no lower than 1.5B RMB, revenue no lower than 200M RMB for last 1year, accumulative R&D expenses no lower than 15% of accumulative revenue for the last 3 years;<br>3. expected market cap no lower than 2B RMB, revenue no lower than 300M RMB for last 1year, accumulative cashflow from operations no lower than 100M RMB for the last 3 years;<br>4. expected market cap no lower than 3B RMB, revenue no lower than 300M RMB for last 1 year;<br>5. expected market cap no lower than 4B RMB, need special approval, with high growth potential and have achieved milestone. |
| Class of shares | Common shares | Common and special voting power shares:<br>1) no. of special voting shares no less than 10% of total no. of shares;<br>2) voting power per special voting share is no more than 10 times of that of common shares;<br>3) no. of common shares no less than 10% of total no. of shares;<br>4) special voting shares cannot be traded, only transferred. | Common and special voting power shares:<br>1) no. of special voting shares no less than 10% of total no. of shares;<br>2) voting power per special voting share is no more than 10 times of that of common shares;<br>3) no. of common shares no less than 10% of total no. of shares;<br>4) special voting shares cannot be traded, only transferred. |
| Investor requirements: accredited investors | Not specified | 100,000RMB average assets 20 days prior to trading, have traded securities for 24 months. | 500,000RMB average assets 20 days prior to trading, have traded securities for 24 months. |